\documentclass{article}

\usepackage{latexsym}

\usepackage{amsfonts}

\begin{document}

\newtheorem{theorem}{Theorem}
\newtheorem{proposition}{Proposition}
\newtheorem{remark}{Remark}
\newtheorem{corollary}{Corollary}
\newtheorem{lemma}{Lemma}
\newtheorem{observation}{Observation}

\newcommand{\qed}{\hfill$\Box$\medskip}

\title{The Complexity of\\ Some Combinatorial Puzzles} 
\author{Holger Petersen\\ 
Reinsburgstr.~75\\
70197 Stuttgart\\ 
}

\maketitle

\begin{abstract}
We show that the decision versions of the
puzzles Knossos and The Hour-Glass are complete for {NP}. 
\end{abstract}

\section{Introduction}

Two puzzles that repeatedly appeared in the Munich newspaper tz are 
Knossos and The Hour-Glass (``Das Stundenglas'')
\cite{tz20100618,tz20100723,tz20101029,tz20110119}, see also \cite{BullsPress2015}.

Knossos asks the reader to insert walls into a grid of cells representing the famous palace of Knossos.
Every room is described by a number that indicates the length of its walls in terms of segments
enclosing the room (we assume that the walls form a connected curve and thus rooms are not 
nested). Note that there is only one shape for rooms with walls of lengths $4$ or $6$, 
where the room with length $6$ contains two cells and can appear in two orientations.
For lengths $8$ and more there are rooms of different shapes containing different numbers 
of cells.

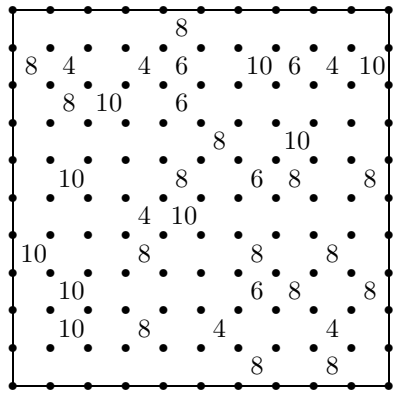
\begin{figure}[ht]
\begin{center}
\setlength{\unitlength}{0.1cm}
\begin{picture}(50,50)
\multiput(0,0)(0,5){11}{\circle*{1}}
\multiput(5,0)(0,5){11}{\circle*{1}}
\multiput(10,0)(0,5){11}{\circle*{1}}
\multiput(15,0)(0,5){11}{\circle*{1}}
\multiput(20,0)(0,5){11}{\circle*{1}}
\multiput(25,0)(0,5){11}{\circle*{1}}
\multiput(30,0)(0,5){11}{\circle*{1}}
\multiput(35,0)(0,5){11}{\circle*{1}}
\multiput(40,0)(0,5){11}{\circle*{1}}
\multiput(45,0)(0,5){11}{\circle*{1}}
\multiput(50,0)(0,5){11}{\circle*{1}}
\multiput(0,0)(5,0){10}{\line(1,0){5}}
\multiput(0,50)(5,0){10}{\line(1,0){5}}
\multiput(0,0)(50,0){2}{\line(0,1){5}}
\multiput(0,5)(50,0){2}{\line(0,1){5}}
\multiput(0,10)(50,0){2}{\line(0,1){5}}
\multiput(0,15)(50,0){2}{\line(0,1){5}}
\multiput(0,20)(50,0){2}{\line(0,1){5}}
\multiput(0,25)(50,0){2}{\line(0,1){5}}
\multiput(0,30)(50,0){2}{\line(0,1){5}}
\multiput(0,35)(50,0){2}{\line(0,1){5}}
\multiput(0,40)(50,0){2}{\line(0,1){5}}
\multiput(0,45)(50,0){2}{\line(0,1){5}}
\put(31,1.5){\,$8$}
\put(41,1.5){\,$8$}
\put(6,6.5){$10$}
\put(16,6.5){\,$8$}
\put(26,6.5){\,$4$}
\put(41,6.5){\,$4$}
\put(6,11.5){$10$}
\put(31,11.5){\,$6$}
\put(36,11.5){\,$8$}
\put(46,11.5){\,$8$}
\put(1,16.5){$10$}
\put(16,16.5){\,$8$}
\put(31,16.5){\,$8$}
\put(41,16.5){\,$8$}
\put(16,21.5){\,$4$}
\put(21,21.5){$10$}
\put(6,26.5){$10$}
\put(21,26.5){\,$8$}
\put(31,26.5){\,$6$}
\put(36,26.5){\,$8$}
\put(46,26.5){\,$8$}
\put(26,31.5){\,$8$}
\put(36,31.5){$10$}
\put(6,36.5){\,$8$}
\put(11,36.5){$10$}
\put(21,36.5){\,$6$}
\put(1,41.5){\,$8$}
\put(6,41.5){\,$4$}
\put(16,41.5){\,$4$}
\put(21,41.5){\,$6$}
\put(31,41.5){$10$}
\put(36,41.5){\,$6$}
\put(41,41.5){\,$4$}
\put(46,41.5){$10$}
\put(21,46.5){\,$8$}
\end{picture}
\end{center}
\caption{Initial grid for Knossos}\label{initial}
\end{figure}

\begin{figure}[ht]
\begin{center}
\setlength{\unitlength}{0.1cm}
\begin{picture}(50,50)
\multiput(0,0)(0,5){11}{\circle*{1}}
\multiput(5,0)(0,5){11}{\circle*{1}}
\multiput(10,0)(0,5){11}{\circle*{1}}
\multiput(15,0)(0,5){11}{\circle*{1}}
\multiput(20,0)(0,5){11}{\circle*{1}}
\multiput(25,0)(0,5){11}{\circle*{1}}
\multiput(30,0)(0,5){11}{\circle*{1}}
\multiput(35,0)(0,5){11}{\circle*{1}}
\multiput(40,0)(0,5){11}{\circle*{1}}
\multiput(45,0)(0,5){11}{\circle*{1}}
\multiput(50,0)(0,5){11}{\circle*{1}}
\multiput(0,0)(5,0){10}{\line(1,0){5}}
\multiput(0,50)(5,0){10}{\line(1,0){5}}
\multiput(0,0)(50,0){2}{\line(0,1){5}}
\multiput(0,5)(50,0){2}{\line(0,1){5}}
\multiput(0,10)(50,0){2}{\line(0,1){5}}
\multiput(0,15)(50,0){2}{\line(0,1){5}}
\multiput(0,20)(50,0){2}{\line(0,1){5}}
\multiput(0,25)(50,0){2}{\line(0,1){5}}
\multiput(0,30)(50,0){2}{\line(0,1){5}}
\multiput(0,35)(50,0){2}{\line(0,1){5}}
\multiput(0,40)(50,0){2}{\line(0,1){5}}
\multiput(0,45)(50,0){2}{\line(0,1){5}}
\put(31,1.5){\,$8$}
\put(41,1.5){\,$8$}
\put(6,6.5){$10$}
\put(16,6.5){\,$8$}
\put(26,6.5){\,$4$}
\put(41,6.5){\,$4$}
\put(6,11.5){$10$}
\put(31,11.5){\,$6$}
\put(36,11.5){\,$8$}
\put(46,11.5){\,$8$}
\put(1,16.5){$10$}
\put(16,16.5){\,$8$}
\put(31,16.5){\,$8$}
\put(41,16.5){\,$8$}
\put(16,21.5){\,$4$}
\put(21,21.5){$10$}
\put(6,26.5){$10$}
\put(21,26.5){\,$8$}
\put(31,26.5){\,$6$}
\put(36,26.5){\,$8$}
\put(46,26.5){\,$8$}
\put(26,31.5){\,$8$}
\put(36,31.5){$10$}
\put(6,36.5){\,$8$}
\put(11,36.5){$10$}
\put(21,36.5){\,$6$}
\put(1,41.5){\,$8$}
\put(6,41.5){\,$4$}
\put(16,41.5){\,$4$}
\put(21,41.5){\,$6$}
\put(31,41.5){$10$}
\put(36,41.5){\,$6$}
\put(41,41.5){\,$4$}
\put(46,41.5){$10$}
\put(21,46.5){\,$8$}
\put(15,0){\line(0,1){5}}
\put(25,0){\line(0,1){5}}
\put(35,0){\line(0,1){5}}
\put(5,5){\line(0,1){5}}
\put(15,5){\line(0,1){5}}
\put(25,5){\line(0,1){5}}
\put(30,5){\line(0,1){5}}
\put(35,5){\line(0,1){5}}
\put(40,5){\line(0,1){5}}
\put(45,5){\line(0,1){5}}
\put(5,10){\line(0,1){5}}
\put(10,10){\line(0,1){5}}
\put(20,10){\line(0,1){5}}
\put(25,10){\line(0,1){5}}
\put(35,10){\line(0,1){5}}
\put(45,10){\line(0,1){5}}
\put(5,15){\line(0,1){5}}
\put(15,15){\line(0,1){5}}
\put(20,15){\line(0,1){5}}
\put(25,15){\line(0,1){5}}
\put(35,15){\line(0,1){5}}
\put(45,15){\line(0,1){5}}
\put(5,20){\line(0,1){5}}
\put(10,20){\line(0,1){5}}
\put(15,20){\line(0,1){5}}
\put(20,20){\line(0,1){5}}
\put(30,20){\line(0,1){5}}
\put(35,20){\line(0,1){5}}
\put(40,20){\line(0,1){5}}
\put(45,20){\line(0,1){5}}
\put(15,25){\line(0,1){5}}
\put(25,25){\line(0,1){5}}
\put(35,25){\line(0,1){5}}
\put(45,25){\line(0,1){5}}
\put(10,30){\line(0,1){5}}
\put(15,30){\line(0,1){5}}
\put(20,30){\line(0,1){5}}
\put(30,30){\line(0,1){5}}
\put(45,30){\line(0,1){5}}
\put(10,35){\line(0,1){5}}
\put(15,35){\line(0,1){5}}
\put(25,35){\line(0,1){5}}
\put(30,35){\line(0,1){5}}
\put(35,35){\line(0,1){5}}
\put(40,35){\line(0,1){5}}
\put(45,35){\line(0,1){5}}
\put(5,40){\line(0,1){5}}
\put(10,40){\line(0,1){5}}
\put(15,40){\line(0,1){5}}
\put(20,40){\line(0,1){5}}
\put(30,40){\line(0,1){5}}
\put(35,40){\line(0,1){5}}
\put(40,40){\line(0,1){5}}
\put(45,40){\line(0,1){5}}
\put(10,45){\line(0,1){5}}
\put(15,45){\line(0,1){5}}
\put(30,45){\line(0,1){5}}
\put(40,45){\line(0,1){5}}
\put(0,5){\line(1,0){5}}
\put(25,5){\line(1,0){5}}
\put(35,5){\line(1,0){5}}
\put(40,5){\line(1,0){5}}
\put(45,5){\line(1,0){5}}
\put(5,10){\line(1,0){5}}
\put(10,10){\line(1,0){5}}
\put(15,10){\line(1,0){5}}
\put(20,10){\line(1,0){5}}
\put(25,10){\line(1,0){5}}
\put(30,10){\line(1,0){5}}
\put(40,10){\line(1,0){5}}
\put(10,15){\line(1,0){5}}
\put(25,15){\line(1,0){5}}
\put(30,15){\line(1,0){5}}
\put(35,15){\line(1,0){5}}
\put(40,15){\line(1,0){5}}
\put(5,20){\line(1,0){5}}
\put(15,20){\line(1,0){5}}
\put(25,20){\line(1,0){5}}
\put(35,20){\line(1,0){5}}
\put(45,20){\line(1,0){5}}
\put(0,25){\line(1,0){5}}
\put(10,25){\line(1,0){5}}
\put(15,25){\line(1,0){5}}
\put(20,25){\line(1,0){5}}
\put(25,25){\line(1,0){5}}
\put(30,25){\line(1,0){5}}
\put(40,25){\line(1,0){5}}
\put(0,30){\line(1,0){5}}
\put(5,30){\line(1,0){5}}
\put(10,30){\line(1,0){5}}
\put(20,30){\line(1,0){5}}
\put(25,30){\line(1,0){5}}
\put(30,30){\line(1,0){5}}
\put(35,30){\line(1,0){5}}
\put(40,30){\line(1,0){5}}
\put(15,35){\line(1,0){5}}
\put(20,35){\line(1,0){5}}
\put(30,35){\line(1,0){5}}
\put(35,35){\line(1,0){5}}
\put(45,35){\line(1,0){5}}
\put(0,40){\line(1,0){5}}
\put(5,40){\line(1,0){5}}
\put(15,40){\line(1,0){5}}
\put(20,40){\line(1,0){5}}
\put(25,40){\line(1,0){5}}
\put(40,40){\line(1,0){5}}
\put(5,45){\line(1,0){5}}
\put(15,45){\line(1,0){5}}
\put(20,45){\line(1,0){5}}
\put(25,45){\line(1,0){5}}
\put(35,45){\line(1,0){5}}
\put(40,45){\line(1,0){5}}
\end{picture}


\end{center}

\caption{Solution for Knossos}\label{solution}
\end{figure}
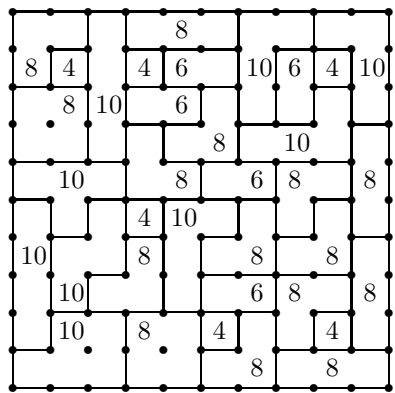

As an example consider the initial grid in Figure~\ref{initial}. 
A possible solution is depicted in Figure~\ref{solution}.

The Hour-Glass consists of two triangular shapes containing numbers and a target sum. The
goal is to find a path from the first row to the last row such that the numbers
traversed add up to the target sum. Each step moves diagonally down the structure.

\begin{figure}[ht]
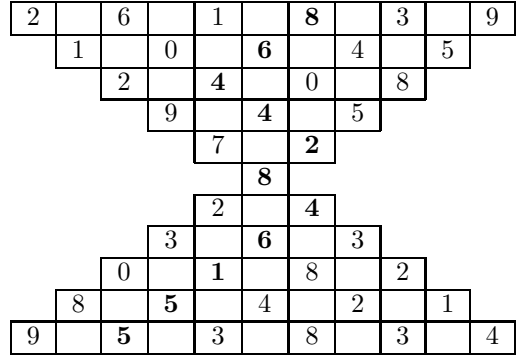

\begin{center}
\begin{tabular}{|c|c|c|c|c|c|c|c|c|c|c|c}\hline
2 &  &  6  & & $ 1 $ & & \bf 8 & & 3 & & 9\\\hline
\multicolumn{1}{c|}{} & 1 &   &   0 &  & \bf 6 &  & 4 & & 5 \\\cline{2-10}
\multicolumn{2}{c|}{} & 2 &  &   \bf 4  &  & 0 &  &  8 \\\cline{3-9}
\multicolumn{3}{c|}{} &   9 &   & \bf 4 &   &  5  \\\cline{4-8}
\multicolumn{4}{c|}{} &   7 &   & \bf 2  \\\cline{5-7}
\multicolumn{5}{c|}{} &  \bf  8    \\\cline{5-7}
\multicolumn{4}{c|}{} &   2 &   &   \bf 4 \\\cline{4-8}
\multicolumn{3}{c|}{} & 3 &  &   \bf  6 &  & 3  \\\cline{3-9}
\multicolumn{2}{c|}{} & 0 &   &   \bf 1 &  & 8 &  &  2 \\\cline{2-10}
\multicolumn{1}{c|}{} & 8 &  &  \bf 5 &  & 4 & & 2 &  & 1 \\\hline
9 &  &   \bf  5  & & 3 & & 8 & & 3 & & 4\\\hline
\end{tabular}
\end{center}
\caption{The Hour-Glass}\label{puzzlehg}
\end{figure}

An example is shown in Figure~\ref{puzzlehg}, where a solution for the target sum
53 is indicated by bold-face numbers.

We generalize these puzzles in several ways:
\begin{enumerate}
\item The original puzzles are of a fixed size (a $10\times 10$-grid for Knossos and a structure with 
11 rows for The Hour-Glass), while we consider arbitrary sizes.
\item As usual for puzzles, in the original versions a solution always exists and the task is to find
it. We consider here also instances that might not have a solution. The problem is to decide 
whether the puzzles have a solution.
\item In the original puzzles the numbers are bounded  (10 for Knossos and  
9 for The Hour-Glass), whereas we allow arbitrary numbers.
\end{enumerate}
The first two generalizations are necessary in order to investigate a decision problem in the
framework of asymptotic complexity. A discussion of the relationship between the 
challenge presented by a puzzle and the complexity of its generalization can be
found in \cite{Eppstein}. 

The last generalization is also essential for The Hour-Glass (under reasonable assumptions) and 
will be discussed separately in Section~\ref{bounds}.

\section{Results}\label{results}

\begin{theorem}
Generalized Knossos is complete in {NP}.
\end{theorem}
{Proof.} 
The problem is easily seen to be in {NP}. Given a grid, 
the positions of walls are guessed and marked. For every
number, the reachable cells can be determined by a variation 
of a recursive method used in computer graphics
for contour filling by connectivity, see \cite[Section~8.4]{Pavlidis82}. 
With the help of this algorithm 
it can be checked that all cells belong to some 
room and that every room contains at most one number.
By starting at one wall and counting the walls of the
circumference it can be verified that the number of walls
is consistent with the inscribed number. Since 
there is a constant upper bound on the number of times a cell is
visited, the complexity is linear in the size of the grid.

For showing hardness we reduce the well-known 
{NP}-complete satisfiability problem for 
boolean formulas (SAT) to Generalized Knossos, see e.g. 
\cite[p.\ 280]{Sipser06}. 

In order to represent the following constructions 
in a clear way, we replace the number $4$ by the 
symbol $\Box$, indicating that the four positions 
surrounding the cell will necessarily contain
walls in every solution. Such cells will serve
as borders and fill space not involved in the construction.

A truth value will in general be represented by
the position of a room with walls of length $6$ containing 
two cells. Two neighboring cells of the cell with inscription
$6$ are empty and the position of the room will indicate
the truth value of a variable.  

The first task we will explain is how to propagate a truth 
value of a variable to 
a position on the grid where it can be processed. We 
describe this for a horizontal propagation. It should be
clear how to generalize the construction of Figure~\ref{signal}
to other directions. In Figure~\ref{corner} we describe 
how to change the direction of propagation. 
 

\newcommand{\signalrow}[8]{
\put(1,#1){#2}
\put(6,#1){#3}
\put(11,#1){#4}
\put(16,#1){#5}
\put(21,#1){#6}
\put(26,#1){#7}
\put(31,#1){#8}
}

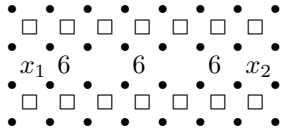
\begin{figure}[ht]
\begin{center}
\setlength{\unitlength}{0.1cm}
\begin{picture}(35,20)
\multiput(0,0)(0,5){4}{\circle*{1}}
\multiput(5,0)(0,5){4}{\circle*{1}}
\multiput(10,0)(0,5){4}{\circle*{1}}
\multiput(15,0)(0,5){4}{\circle*{1}}
\multiput(20,0)(0,5){4}{\circle*{1}}
\multiput(25,0)(0,5){4}{\circle*{1}}
\multiput(30,0)(0,5){4}{\circle*{1}}
\multiput(35,0)(0,5){4}{\circle*{1}}

\signalrow{1.5}{$\Box$}{$\Box$}{$\Box$}{$\Box$}{$\Box$}{$\Box$}{$\Box$}
\signalrow{6.5}{$x_1$}{$6$}{}{$6$}{}{$6$}{$x_2$}
\signalrow{11.5}{$\Box$}{$\Box$}{$\Box$}{$\Box$}{$\Box$}{$\Box$}{$\Box$}

\end{picture}
\end{center}
\caption{Propagation of truth value}\label{signal}
\end{figure}



\newcommand{\cornerrow}[4]{
\put(1,#1){#2}
\put(6,#1){#3}
\put(11,#1){#4}
}

\begin{figure}[ht]
\begin{center}
\setlength{\unitlength}{0.1cm}
\begin{picture}(15,20)
\multiput(0,0)(0,5){4}{\circle*{1}}
\multiput(5,0)(0,5){4}{\circle*{1}}
\multiput(10,0)(0,5){4}{\circle*{1}}
\multiput(15,0)(0,5){4}{\circle*{1}}

\cornerrow{1.5}{$\Box$}{$x_2$}{$\Box$}
\cornerrow{6.5}{$x_1$}{$6$}{$\Box$}
\cornerrow{11.5}{$\Box$}{$\Box$}{$\Box$}
\end{picture}
\end{center}
\caption{Corner}\label{corner}
\end{figure}


Notice that depending on the distance between 
the origin of a propagated value and its destination,
it might arrive ``out of phase'' or we might have
to invert the propagated value. We can achieve these
goals by inserting a ``phase-shift'' construction
that contains a room with $8$ walls instead of $6$.
This is outlined in Figure~\ref{shift}.


\newcommand{\shiftrow}[9]{
\put(1,#1){#2}
\put(6,#1){#3}
\put(11,#1){#4}
\put(16,#1){#5}
\put(21,#1){#6}
\put(26,#1){#7}
\put(31,#1){#8}
\put(36,#1){#9}
}

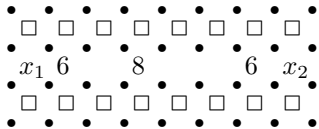
\begin{figure}[ht]
\begin{center}
\setlength{\unitlength}{0.1cm}
\begin{picture}(40,20)
\multiput(0,0)(0,5){4}{\circle*{1}}
\multiput(5,0)(0,5){4}{\circle*{1}}
\multiput(10,0)(0,5){4}{\circle*{1}}
\multiput(15,0)(0,5){4}{\circle*{1}}
\multiput(20,0)(0,5){4}{\circle*{1}}
\multiput(25,0)(0,5){4}{\circle*{1}}
\multiput(30,0)(0,5){4}{\circle*{1}}
\multiput(35,0)(0,5){4}{\circle*{1}}
\multiput(40,0)(0,5){4}{\circle*{1}}

\shiftrow{1.5}{$\Box$}{$\Box$}{$\Box$}{$\Box$}{$\Box$}{$\Box$}{$\Box$}{$\Box$}
\shiftrow{6.5}{$x_1$}{$6$}{}{$8$}{}{}{$6$}{$x_2$}
\shiftrow{11.5}{$\Box$}{$\Box$}{$\Box$}{$\Box$}{$\Box$}{$\Box$}{$\Box$}{$\Box$}

\end{picture}
\end{center}
\caption{Phase shift}\label{shift}
\end{figure}


Given a boolean formula, it is necessary to encode 
a truth-assignment in terms of distribution of walls. In 
Figure~\ref{choice} we demonstrate how to do this for
a variable $x$ that occurs at most 6 times in literals of the 
formula to be encoded. Three of the copies are initially complemented
but can be  inverted by the  ``phase-shift''-construction presented
above.
It is obvious how to generalize the
construction to larger numbers (see also Section~\ref{bounds}).

We now outline, why the construction correctly assigns truth 
values to the copies of variable $x$. Notice that not
both of the cells $A$ and $B$ can be in a room including the
central $12$, since this would require at least $20$ walls.
At least one of the cells is not in the room with $26$ walls.
Thus either all cells above the cells $x_1$ through $x_3$
or $\bar x_1$ through $\bar x_3$ are in one room with $12$ walls.
This induces a consistent assignment of the copies of $x$
($x_1$, $x_2$, and $x_3$) and of the copies of $\bar x$
($\bar x_1$, $\bar x_2$, and $\bar x_3$).


\newcommand{\choicerowl}[8]{
\put(1,#1){#2}
\put(6,#1){#3}
\put(11,#1){#4}
\put(16,#1){#5}
\put(21,#1){#6}
\put(26,#1){#7}
\put(31,#1){#8}

}
\newcommand{\choicerowr}[7]{
\put(36,#1){#2}
\put(41,#1){#3}
\put(46,#1){#4}
\put(51,#1){#5}
\put(56,#1){#6}
\put(61,#1){#7}
}

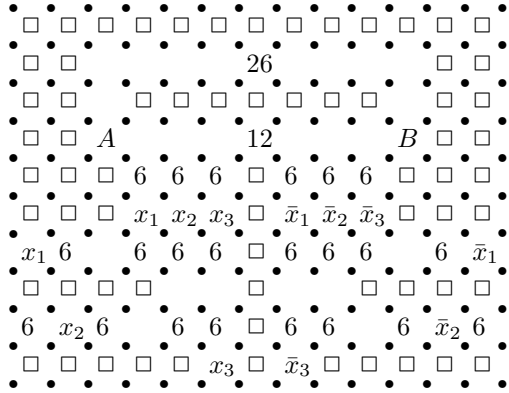
\begin{figure}[ht]
\begin{center}
\setlength{\unitlength}{0.1cm}
\begin{picture}(65,50)
\multiput(0,0)(0,5){11}{\circle*{1}}
\multiput(5,0)(0,5){11}{\circle*{1}}
\multiput(10,0)(0,5){11}{\circle*{1}}
\multiput(15,0)(0,5){11}{\circle*{1}}
\multiput(20,0)(0,5){11}{\circle*{1}}
\multiput(25,0)(0,5){11}{\circle*{1}}
\multiput(30,0)(0,5){11}{\circle*{1}}
\multiput(35,0)(0,5){11}{\circle*{1}}
\multiput(40,0)(0,5){11}{\circle*{1}}
\multiput(45,0)(0,5){11}{\circle*{1}}
\multiput(50,0)(0,5){11}{\circle*{1}}
\multiput(55,0)(0,5){11}{\circle*{1}}
\multiput(60,0)(0,5){11}{\circle*{1}}
\multiput(65,0)(0,5){11}{\circle*{1}}
\choicerowl{1.5}{$\Box$}{$\Box$}{$\Box$}{$\Box$}{$\Box$}{$x_3$}{$\Box$}
\choicerowr{1.5}{$\bar x_3$}{$\Box$}{$\Box$}{$\Box$}{$\Box$}{$\Box$}
\choicerowl{6.5}{$6$}{$x_2$}{$6$}{}{$6$}{$6$}{$\Box$}
\choicerowr{6.5}{$6$}{$6$}{}{$6$}{$\bar x_2$}{$6$}
\choicerowl{11.5}{$\Box$}{$\Box$}{$\Box$}{$\Box$}{}{}{$\Box$}
\choicerowr{11.5}{}{}{$\Box$}{$\Box$}{$\Box$}{$\Box$}
\choicerowl{16.5}{$x_1$}{$6$}{}{$6$}{$6$}{$6$}{$\Box$}
\choicerowr{16.5}{$6$}{$6$}{$6$}{}{$6$}{$\bar x_1$}
\choicerowl{21.5}{$\Box$}{$\Box$}{$\Box$}{$x_1$}{$x_2$}{$x_3$}{$\Box$}
\choicerowr{21.5}{$\bar x_1$}{$\bar x_2$}{$\bar x_3$}{$\Box$}{$\Box$}{$\Box$}
\choicerowl{26.5}{$\Box$}{$\Box$}{$\Box$}{$6$}{$6$}{$6$}{$\Box$}
\choicerowr{26.5}{$6$}{$6$}{$6$}{$\Box$}{$\Box$}{$\Box$}
\choicerowl{31.5}{$\Box$}{$\Box$}{$A$}{}{}{}{$12$}
\choicerowr{31.5}{}{}{}{$B$}{$\Box$}{$\Box$}
\choicerowl{36.5}{$\Box$}{$\Box$}{}{$\Box$}{$\Box$}{$\Box$}{$\Box$}
\choicerowr{36.5}{$\Box$}{$\Box$}{$\Box$}{}{$\Box$}{$\Box$}
\choicerowl{41.5}{$\Box$}{$\Box$}{}{}{}{}{$26$}
\choicerowr{41.5}{}{}{}{}{$\Box$}{$\Box$}
\choicerowl{46.5}{$\Box$}{$\Box$}{$\Box$}{$\Box$}{$\Box$}{$\Box$}{$\Box$}
\choicerowr{46.5}{$\Box$}{$\Box$}{$\Box$}{$\Box$}{$\Box$}{$\Box$}
\end{picture}
\caption{Construction for choice of truth values}\label{choice}
\end{center}
\end{figure}


We now know how to assign values to variables and how to 
propagate these values. But the encoding on a grid 
dictates a planar layout, and we may have to introduce cross-overs
into the construction. Figure~\ref{crossover} shows how to do this.


\newcommand{\crossrowl}[8]{
\put(1,#1){#2}
\put(6,#1){#3}
\put(11,#1){#4}
\put(16,#1){#5}
\put(21,#1){#6}
\put(26,#1){#7}
\put(31,#1){#8}
}

\newcommand{\crossrowr}[5]{
\put(36,#1){#2}
\put(41,#1){#3}
\put(46,#1){#4}
\put(51,#1){#5}
}

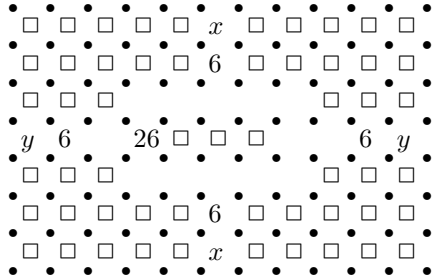
\begin{figure}[ht]
\begin{center}
\setlength{\unitlength}{0.1cm}
\begin{picture}(55,40)
\multiput(0,0)(0,5){8}{\circle*{1}}
\multiput(5,0)(0,5){8}{\circle*{1}}
\multiput(10,0)(0,5){8}{\circle*{1}}
\multiput(15,0)(0,5){8}{\circle*{1}}
\multiput(20,0)(0,5){8}{\circle*{1}}
\multiput(25,0)(0,5){8}{\circle*{1}}
\multiput(30,0)(0,5){8}{\circle*{1}}
\multiput(35,0)(0,5){8}{\circle*{1}}
\multiput(40,0)(0,5){8}{\circle*{1}}
\multiput(45,0)(0,5){8}{\circle*{1}}
\multiput(50,0)(0,5){8}{\circle*{1}}
\multiput(55,0)(0,5){8}{\circle*{1}}

\crossrowl{1.5}{$\Box$}{$\Box$}{$\Box$}{$\Box$}{$\Box$}{$x$}{$\Box$}
\crossrowr{1.5}{$\Box$}{$\Box$}{$\Box$}{$\Box$}
\crossrowl{6.5}{$\Box$}{$\Box$}{$\Box$}{$\Box$}{$\Box$}{$6$}{$\Box$}
\crossrowr{6.5}{$\Box$}{$\Box$}{$\Box$}{$\Box$}
\crossrowl{11.5}{$\Box$}{$\Box$}{$\Box$}{}{}{}{}
\crossrowr{11.5}{}{$\Box$}{$\Box$}{$\Box$}
\crossrowl{16.5}{$y$}{$6$}{}{$26$}{$\Box$}{$\Box$}{$\Box$}
\crossrowr{16.5}{}{}{$6$}{$y$}
\crossrowl{21.5}{$\Box$}{$\Box$}{$\Box$}{}{}{}{}
\crossrowr{21.5}{}{$\Box$}{$\Box$}{$\Box$}
\crossrowl{26.5}{$\Box$}{$\Box$}{$\Box$}{$\Box$}{$\Box$}{$6$}{$\Box$}
\crossrowr{26.5}{$\Box$}{$\Box$}{$\Box$}{$\Box$}
\crossrowl{31.5}{$\Box$}{$\Box$}{$\Box$}{$\Box$}{$\Box$}{$x$}{$\Box$}
\crossrowr{31.5}{$\Box$}{$\Box$}{$\Box$}{$\Box$}
\end{picture}
\end{center}
\caption{Construction for cross-over}\label{crossover}
\end{figure}

We first argue, that the value of variable $x$ is transferred
correctly in the sense, that the upper $6$ and the cell
marked $x$ are in one room, if and only if the lower $6$
is not in one room with the cell containing $x$. 

If both
rooms including the cells with $6$ do not contain $x$, then
the room including the $26$ can have at most $14$ walls
and therefore this option is excluded. If on the other
hand both rooms contain $x$, then the central three cells
are isolated and the room including the $26$ cannot have
connected walls. This show correctness for $x$.

Without loss of generality we now assume that the room 
with the upper $6$ does not contain the cell
marked $x$. If both rooms containing the
left and right $6$ also include the cells
marked $y$, then the wall of the
room including $26$ has length $28$.
If If both rooms do not include the cells
marked $y$, then the length is $24$. 
This shows that $y$ is transferred correctly.

Next we show how to compute the AND of values of two
variables $x$ and $y$. We note that together with the 
option to invert values (see Figure~\ref{shift}) all
boolean functions can be computed. 


\begin{figure}[ht]
\begin{center}
\setlength{\unitlength}{0.1cm}
\begin{picture}(25,35)
\multiput(5,0)(0,5){7}{\circle*{1}}
\multiput(10,0)(0,5){7}{\circle*{1}}
\multiput(15,0)(0,5){7}{\circle*{1}}
\multiput(20,0)(0,5){7}{\circle*{1}}
\multiput(25,0)(0,5){7}{\circle*{1}}

\put(6,1.5){$\Box$}
\put(11,1.5){\,$z$}
\put(16,1.5){$\Box$}
\put(21,1.5){$\Box$}
\put(6,6.5){$\Box$}
\put(11,6.5){$10$}
\put(16,6.5){$\Box$}
\put(21,6.5){$\Box$}
\put(6,11.5){$\Box$}
\put(21,11.5){$\Box$}
\put(6,16.5){$\Box$}
\put(21,16.5){$\Box$}
\put(6,21.5){$\Box$}
\put(11,21.5){\,$6$}
\put(16,21.5){\,$6$}
\put(21,21.5){$\Box$}
\put(6,26.5){$\Box$}
\put(11,26.5){\,$x$}
\put(16,26.5){\,$y$}
\put(21,26.5){$\Box$}
\put(26,31.5){}
\put(36,31.5){}

\end{picture}
\end{center}
\caption{AND}\label{and}
\end{figure}

The value of result variable $z$ depends
on the conjunction of $x$ and $y$.

\begin{figure}[ht]
\begin{center}
\setlength{\unitlength}{0.1cm}
\begin{picture}(25,30)
\multiput(5,0)(0,5){6}{\circle*{1}}
\multiput(10,0)(0,5){6}{\circle*{1}}
\multiput(15,0)(0,5){6}{\circle*{1}}
\multiput(20,0)(0,5){6}{\circle*{1}}
\multiput(25,0)(0,5){6}{\circle*{1}}

\multiput(6,1.5)(5,0){4}{$\Box$}
\put(6,6.5){$\Box$}
\put(16,6.5){$8$}
\put(21,6.5){$\Box$}
\put(6,11.5){$\Box$}
\put(11,11.5){$A$}
\put(21,11.5){$\Box$}
\put(6,16.5){$\Box$}
\put(11,16.5){$6$}
\put(16,16.5){$\Box$}
\put(21,16.5){$\Box$}
\put(6,21.5){$\Box$}
\put(11,21.5){$x$}
\put(16,21.5){$\Box$}
\put(21,21.5){$\Box$}
\end{picture}
\end{center}
\caption{Slack construction}\label{slack}
\end{figure}
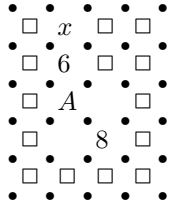

We also have to ``erase'' values if an odd
number of values is required, since by the 
construction in Figure~\ref{choice} an even
number of values is generated. We do this
with the help of the ``slack'' construction
described in Figure~\ref{slack}.  

\bigskip


Finally the resulting value of the boolean 
formula encoded has to be true. This can be
guaranteed by closing the propagating construction 
of Figure~\ref{signal}.\qed

\begin{theorem}
Generalized Hour-Glass is complete in {NP}.
\end{theorem}
{Proof.}  The problem is in {NP} by the following algorithm.
A path from the first to the last row of the input
is guessed and the binary numbers along the path are added.
The sum is compared to the target and the input is accepted
if the values are equal.

For hardness we reduce the {NP}-complete 
problem SUBSET-SUM to The Hour-Glass,
see \cite[Theorem~7.56]{Sipser06}.

The input of SUBSET-SUM consists of a sequence $S = (x_1, \ldots,  x_n)$ 
of positive integers and a positive integer $t$, where the integers are 
encoded in binary. There is a subset $S' \subseteq \{1, \ldots, n \}$ with
$\sum_{i \in S'} x_i = t$.

Without loss of generality $n$ is even, since we can add 0 to a list
of odd length without affecting possible sums.

For a sequence $S = (x_1, \ldots,  x_n)$  and target $t$ 
we form a structure with
$2n-1$ rows and put $x_i$ into column $n-1$ in row $2i-1$. All 
other non-empty positions contain 0. As an example we include the structure
for $n = 6$:

\begin{center}
\begin{tabular}{|c|c|c|c|c|c|c|c|c|c|c|c}\hline
0 &  & 0  & & $x_1$ & & 0 & & 0 & & 0\\\hline
\multicolumn{1}{c|}{} & 0 &   & 0 &  & 0 &  & 0 & & 0 \\\cline{2-10}
\multicolumn{2}{c|}{} & 0 &  & $x_2$ &  & 0 &  &  0 \\\cline{3-9}
\multicolumn{3}{c|}{} & 0 &   & 0 &   &  0  \\\cline{4-8}
\multicolumn{4}{c|}{} & $x_3$ &   &  0  \\\cline{5-7}
\multicolumn{5}{c|}{} & 0    \\\cline{5-7}
\multicolumn{4}{c|}{} & $x_4$ & 0  &  0 \\\cline{4-8}
\multicolumn{3}{c|}{} & 0 &  & 0 &  & 0  \\\cline{3-9}
\multicolumn{2}{c|}{} & 0 &   & $x_5$ &  & 0 &  &  0 \\\cline{2-10}
\multicolumn{1}{c|}{} & 0 &  & 0 &  & 0 & & 0 &  & 0 \\\hline
0 &  & 0  & & $x_6$ & & 0 & & 0 & & 0\\\hline
\end{tabular}
\end{center}

Suppose there is a $S' \subseteq \{1, \ldots, n \}$ solving 
the instance of SUBSET-SUM.
We construct a path along cells adding up to $t$.
If $1 \in S'$ we choose the number in column $n-1$ in row 1 as the
first number of the path, otherwise we choose the 0 in column $n+1$.
If the path has been constructed up to row $2i -1$ for $i \ge 1$,
the next cell is the 0 in column $n$. If the path has been 
constructed up to row $2i $, the next step is to $x_i$ in column $n-1$
if $i \in S'$ and to the 0 in column $n+1$ if $i \not\in S'$. 
Since $\sum_{i \in S'} x_i = t$, the numbers along the path
add up to $t$. If conversely a path adds up to $t$, we can
construct a $S'$ by collecting all rows in which numbers in
column $n-1$ are on the path. \qed

\section{Bounds on Numbers}\label{bounds}
If in Knossos only 4 is allowed as a number of walls, then a solution exists if all 
cells contain this number. This can be checked quickly.

In the construction from Section~\ref{results} only the choice of 
a truth value requires a variable size of rooms. In order to bound the
size of the rooms, we duplicate the structure for the choice and 
connect the structures by signals. If the structures for choice are
restricted to four output values and the upper room (with size 26)
is split into a chain of rooms of size 6, then the bound 10 can be 
satisfied.
The bound 26 in the cross-over
structure can likely be improved, but whether the construction is 
possible with bound 10 remains open. 

In the case of The Hour-Glass any constant bound on
the numbers in the structure admits a non-deterministic solution
in logarithmic space. The $O(n)$ numbers along a path are added
in $O(\log n)$ bits and compared to the target sum. Unless 
{NP} and {NL} coincide, the constant bound makes the 
problem easier.


\end{document}